\documentclass[runningheads]{llncs}
\usepackage{graphicx}
\usepackage{diagbox}
\usepackage{xcolor}
\usepackage{todonotes}

\begin{document}
\title{The Impact of Domain Shift on Left and Right Ventricle Segmentation in Short Axis Cardiac MR Images}
\titlerunning{Impact of Domain Shift on LV, RV Segmentation in CMR}
\author{Devran Ugurlu* \inst{1} \and 
Esther Puyol-Ant\'on \inst{1}  \and
Bram Ruijsink \inst{1,4} \and
Alistair Young \inst{1} \and 
In\^{e}s Machado \inst{1} \and 
Kerstin Hammernik \inst{2,3} \and
Andrew P. King \inst{1} \and
Julia A. Schnabel \inst{1,2,5} 
} 
\authorrunning{Devran Ugurlu et al.}

\institute{School of Biomedical Engineering \& Imaging Sciences, King's College London, UK \and Technical University of Munich, Germany \and Biomedical Image Analysis Group, Imperial College London, UK \and Department of Adult and Paediatric Cardiology, Guy’s and St Thomas’ NHS Foundation Trust, London, UK \and Helmholtz Center Munich, Germany}
\maketitle              % typeset the header of the contribution
\begin{abstract}
Domain shift refers to the difference in the data distribution of two datasets, normally between the training set and the test set for machine learning algorithms. Domain shift is a serious problem for generalization of machine learning models and it is well-established that a domain shift between the training and test sets may cause a drastic drop in the model's performance. In medical imaging, there can be many sources of domain shift such as different scanners or scan protocols, different pathologies in the patient population, anatomical differences in the patient population (e.g. men vs women) etc. Therefore, in order to train models that have good generalization performance, it is important to be aware of the domain shift problem, its potential causes and to devise ways to address it. In this paper, we study the effect of domain shift on left and right ventricle blood pool segmentation in short axis cardiac MR images. Our dataset contains short axis images from 4 different MR scanners and 3 different pathology groups. The training is performed with nnUNet. The results show that scanner differences cause a greater drop in performance compared to changing the pathology group, and that the impact of domain shift is greater on right ventricle segmentation compared to left ventricle segmentation. Increasing the number of training subjects increased cross-scanner performance more than in-scanner performance at small training set sizes, but this difference in improvement decreased with larger training set sizes. Training models using data from multiple scanners improved cross-domain performance.

\end{abstract}
\section{Introduction}
Cardiac magnetic resonance (CMR) allows non-invasive and radiation-free imaging of the heart and is the recommended imaging modality for many clinical scenarios \cite{vonKnobelsdorff-Brenkenhoff2017}. Cine CMR involves acquiring images at several time steps through the cardiac cycle. These images can be used to measure diagnostic metrics such as left and right ventricle (LV and RV) volumes and ejection fraction (EF) \cite{10.1007/978-3-030-68107-4_10,RUIJSINK2020684}. In order to calculate these metrics, segmentation of the LV and RV from the cine images is required. Manual segmentation requires expertise, is very time-consuming and is prone to operator bias and error. Hence, automatic segmentation methods are highly desirable. Several challenges have been organized to tackle this problem and many methods have been proposed. Recent challenges such as ACDC \cite{8360453} and M\&Ms \cite{9458279} have been dominated by U-Net-based \cite{UNet} deep learning methods, with nnUNet \cite{Isensee2021} being the best overall performer in the most recent M\&Ms challenge. 

A significant problem with deep learning methods is that their performance can drastically drop if the data distribution of the training set and test set are different, and this phenomenon is known as domain shift or dataset shift \cite{10.5555/1462129}. Medical images might have several sources of domain shift such as different scanners or scan protocols, different pathologies in the patient population and anatomical differences in the patient population (e.g. men vs women). Hence, it is recommended for clinicians using AI assistance to be aware of potential sources of domain shift that could cause the AI system to fail \cite{doi:10.1056/NEJMc2104626}.

With intra-domain performance reaching comparable quality to manual expert segmentations \cite{8360453}, analysis of domain shift and ways to improve cross-domain generalization are becoming popular topics in recent studies. Significant cross-scanner performance drop has indeed been observed on the M\&Ms challenge data \cite{9458279} and several studies have proposed various methods to improve cross-domain generalization. On the M\&Ms dataset, an aggressive data augmentation scheme used with nnUNet \cite{10.1007/978-3-030-68107-4_24}, and an approach based on domain-adversarial learning \cite{10.1007/978-3-030-68107-4_23} were shown to improve cross-scanner performance. In another study, an adverserial domain adaptation approach was utilized on a multi-modality dataset showing improved cross-modality performance \cite{10.1007/978-3-030-39074-7_27}. However, there has been relatively little work on investigating and quantifying the relative impact of different sources of domain shift.

In this paper, we study the effect of domain shift on LV and RV blood pool segmentation in end-diastolic (ED) and end-systolic (ES) frames of short axis
cine CMR images. Our dataset contains short axis images from 4 different MR scanners and 3 different pathology groups. The training is performed using the nnUNet model \cite{Isensee2021}.
The contributions of our study are as follows: (1) We separate domains not only by scanner but also by pathology group in order to systematically
analyse the pathology-based and scanner-based domain shifts.
(2) Using a large dataset of 1373 subjects with manual segmentations, we
analyse the impact of training set size on domain shift both on intra-scanner and cross-scanner test sets.

\section{Materials and Methods}
\subsection{Dataset}
We employ a dataset of 1373 cine CMR examinations acquired for clinical purposes from the clinical imaging system of Guy’s and St Thomas’ NHS Foundation Trust Hospital. All patients gave informed consent for research use of their imaging data. For each subject, an experienced cardiologist manually segmented the LV and RV blood pools in the ES and ED frames of the short axis sequence. The scanners and numbers of subjects from each scanner are as follows: Siemens Aera 1.5 T (555 subjects), Philips Ingenia 1.5 T (324 subjects), Philips Achieva 3 T (452 subjects) and Siemens Biograph mMR 3 T (42 subjects). For brevity, we name these domains as A, B, C and D, respectively.

The dataset is partially annotated with the following pathology groups: healthy, congenital heart disease, dilated cardiomyopathy (DCM), hypertrophic cardiomyopathy (HCM), hypertensive cardiomyopathy (CMP), ischaemic, mixed CMP, non-ischaemic CMP, other CMP, other (for example, myocarditis, infiltrative myocardial diseases, etc.). From these, the following groups were created as the pathology domains for our experiments: Pathology Group 1 (PG 1): Healthy, PG 2: DCM, hypertensive CMP, mixed CMP, non-ischaemic CMP. PG 3: HCM. Other groups were removed either due to very small numbers of subjects or high structural variance within the group. DCM and various CMPs were grouped together in PG 2 due to structural similarity. Using the scanners A, B, C and the 3 pathology domains, we created 9 domains
containing different combinations of scanner/pathology. The domain names, descriptions and numbers of subjects are summarized in Table~\ref{tab_domains}.

\begin{table}
\centering
\caption{Description of domains defined within our dataset.}\label{tab_domains}
\begin{tabular}{|l|l|l|l|}
\hline
Single-scanner domains & Scanner & Pathology Group & No. Subjects \\
\hline
A & Siemens Aera 1.5 T & All & 554\\
B & Philips Ingenia 1.5 T & All & 324\\
C & Philips Achieva 3 T & All & 452\\
D & Siemens Biograph\_mMR 3 T & All & 42\\
A1 & Siemens Aera 1.5 T & Healthy & 74\\
A2 & Siemens Aera 1.5 T & DCM, CMP & 37\\
A3 & Siemens Aera 1.5 T & HCM & 15\\
B1 & Philips Ingenia 1.5 T & Healthy & 42\\
B2 & Philips Ingenia 1.5 T & DCM, CMP & 14\\
B3 & Philips Ingenia 1.5 T & HCM & 9\\
C1 & Philips Achieva 3 T & Healthy & 64\\
C2 & Philips Achieva 3 T & DCM, CMP & 36\\
C3 & Philips Achieva 3 T & HCM & 8\\
\hline
\hline
Mixed domains & \multicolumn{3}{|l|}{Domains Included} \\
\hline
AM & \multicolumn{3}{|l|}{A1, A2} \\
CM & \multicolumn{3}{|l|}{C1, C2} \\
1M & \multicolumn{3}{|l|}{A1, B1, C1} \\
2M & \multicolumn{3}{|l|}{A2, C2} \\
M & \multicolumn{3}{|l|}{A, B, C} \\
\hline
\end{tabular}
\end{table}

\subsection{Experimental Setup and Training}
The first set of experiments was aimed at analysing the pathology-based and scanner-based domain shifts separately. From the 9 scanner and pathology-specific domains, the ones that had more than 30 subjects, i.e. A1, A2, B1, C1 and C2
were separated into training and test sets, with 30 subjects in the training set
for each domain. The rest of the domains were only used for testing. In addition, mixed domains were created by mixing the pathologies and scanners and named as follows: AM: A1 + A2, CM: C1 + C2, 1M: A1 + B1 + C1, 2M: A2 + C2. The mixed domain names and descriptions are summarized in Table~\ref{tab_domains}. Training set sizes were also fixed to 30 for mixed domains and subjects were randomly selected with equal proportion from the domains that make up the mixed domains. For example, domain 1M's training set consisted of 10 subjects each from A1, B1 and C1 domains and domain AM's training set consisted of 15 subjects each from domains A1 and A2. This was done to keep the training set size fixed across all domains in order to separate the effect of domain shift on performance from increasing training set size.

The second set of experiments was aimed at measuring the effect of increasing training set size on intra-domain and cross-domain performance. For these experiments, we only analyse the effect of scanner-based domain shift on the A, B, C and D domains because our pathology-specific domains were not large enough for this purpose. Models were trained on A, B, C and M domains, where M is the union of A, B and C, and were trained with 30, 60, 120 and 240 training set sizes. A final model was trained with a training set size of 1064 for domain M. Domain D was only used as a test set for cross-domain testing. 

For all experiments, 2D training was performed using the nnUNet \cite{Isensee2021} model with batch size 8, 250 max epochs and 4 fold cross-validation. We did not use 3D training because of training time constraints. All other parameters, including data augmentation settings, are default settings for the version 2 trainer of nnUNet (nnUNetTrainerV2). An ensemble automatically created by nnUNet from the 4 folds was used for testing.

\section{Results and Discussion}
The mean Dice scores between manual segmentations and nnUNet-produced
segmentations over LV, RV, ED and ES are given in Table~\ref{tab_all_dices}. An interesting observation is that the cross-domain performance is clearly not symmetric. For example, the model trained on domain C2 performs very poorly on domain A1 but the reverse is not true. 

An in-domain dominance is not immediately obvious from Table~\ref{tab_all_dices}. On closer inspection, however, it can be noticed that the best model for every test domain was trained on a domain that includes the same scanner. Furthermore, there are a number of
very poor performances on cross-scanner tests (e.g. train C2, test A1) which is not the case for any in-scanner test.
To test the hypothesis that in-domain performance is better than cross-domain performance, we computed means and standard deviations of Dice scores for intra-scanner, cross-scanner, intra-pathology and cross-pathology cases using the domains A1, A2, A3, B1, B2, B3, C1, C2 and C3, and statistically compared these using Mann-Whitney U tests. The results are given in Table~\ref{tab_in_cross_dices}, which shows a clear effect of domain shift across scanners for both LV and RV and for both ED and ES frames, whereas the domain shift effect is not significant across pathology groups for any segmented structure and frame combination. Furthermore, Table~\ref{tab_in_cross_dices} also shows a more significant cross-domain performance drop on RV segmentations compared to LV segmentations across scanners.

\begin{figure}
\centering
\includegraphics[width=\textwidth]{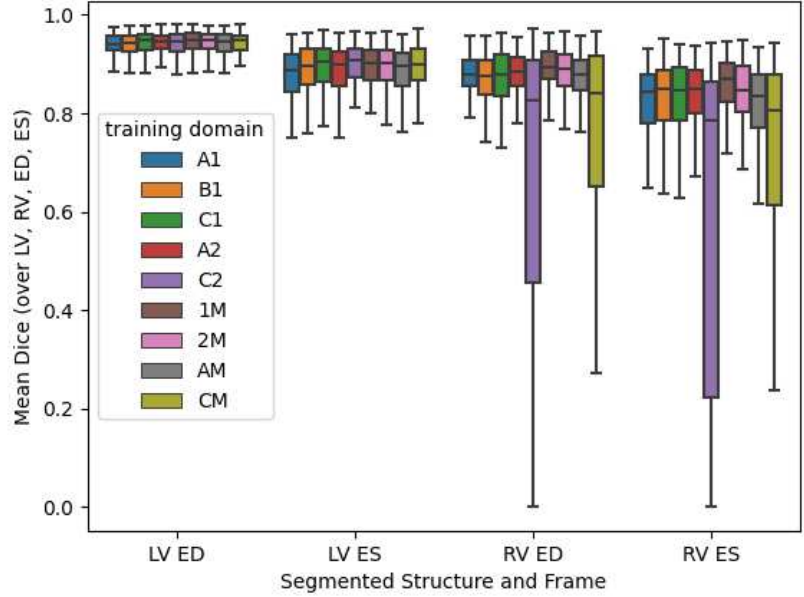}
\caption{Boxplot of Dice scores for different training domains (including mixed pathology domains AM and CM, and mixed scanner domains 1M and 2M) for a single aggregated test domain that combines all of A1, A2, A3, B1, B2, B3, C1, C2, C3. See Table~\ref{tab_domains} for domain descriptions.} \label{fig_all_test_domains_dice_boxplot}
\end{figure}

Examining the significance results in Table~\ref{tab_in_cross_dices}, it can be hypothesized that mixing different scanner domains will improve generalization performance. To investigate this hypothesis, we produced
a boxplot of Dice scores for different training domains (including mixed pathology domains AM and CM, and mixed scanner domains 1M and 2M) for a single pooled test domain that aggregates all of the individual test domains A1, A2, A3, B1, B2, B3, C1, C2 and C3 (see Fig.~\ref{fig_all_test_domains_dice_boxplot}). Note that the numbers of subjects in the training sets were kept fixed when mixing domains in order to remove the effect of increasing training set size from the analysis. Indeed, the healthy group with mixed scanners (1M) had the best overall performance, especially due to its performance on the more difficult RV segmentation task.

\begin{figure}
\centering
\includegraphics[width=\textwidth]{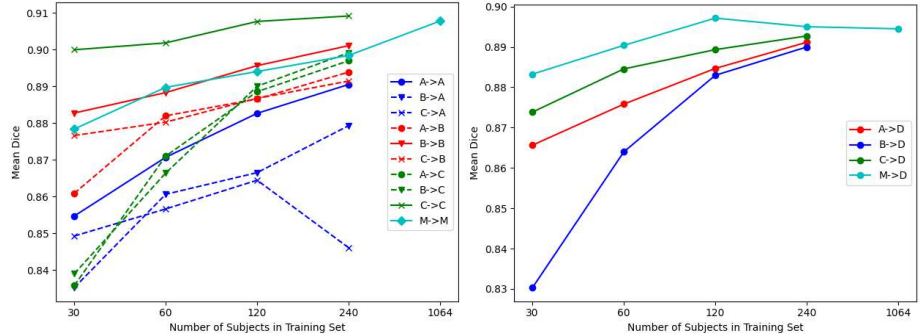}
\caption{Lineplots of mean Dice scores (over LV, RV, ED and ES)
for each domain model to each domain test set with increasing training set size. A$\rightarrow$B means the model was trained on domain A and tested on domain B. On the left lineplot, intra-scanner tests are shown as solid lines and cross-scanner tests as dotted lines.
} \label{fig_training_size}
\end{figure}

In Fig.~\ref{fig_training_size}, we plot the mean Dice scores (over LV, RV, ED and ES) vs. the number of subjects in the training set (30, 60, 120, 240) for each model trained on domains A, B and C (i.e. different scanners) to each test domain A, B, C and D. A model trained on the mixed domain M and tested on domain M and D is also added to the plot. For the mixed domain experiment, an additional training size of 1064 that makes use of all available data is added. It can be seen that for every combination except C$\rightarrow$A, that is, the model trained on domain C and tested on domain A, and M$\rightarrow$D, performance increases with increasing number of subjects in the training set. On C$\rightarrow$A and M$\rightarrow$D, the performance kept increasing until 120 training subjects but dropped when the training size was increased further. These two might be outlier cases since a similar drop was not observed in any of the other tests. For all test set sizes, the model trained on multiple scanners (i.e. M) has comparable performance to that of single-scanner models on intra-scanner experiments and performed better than single scanner models on cross-scanner experiments. In addition, when trained with all available data (1064 subjects), this model's performance matched the best performance of any of the intra-domain models, demonstrating the utility of using large training sets from multiple scanners.
It also generalised well to the unseen domain D, as can be seen from the right-hand plot in Fig.~\ref{fig_training_size}.

\begin{table}
\centering
\caption{Mean Dice scores between manual segmentations and nnUNet-produced
segmentations over LV, RV, ED and ES. The mean Dice score for the best performing model for each test domain is displayed in bold and the second best model is displayed in italic. Multiple scores may be in bold or italic when there are ties. }\label{tab_all_dices}
\resizebox{\columnwidth}{!}{
\begin{tabular}{|c|c|c|c|c|c|c|c|c|c|}
\hline
 \diagbox[width=7em]{Train \\ Domain}{Test \\ Domain} & A1  & B1  & C1 & A2 & C2 & B2 & A3 & B3 & C3 \\
\hline
A1 & 0.866 & 0.859 & 0.873 & 0.901 & 0.872 & 0.890 & 0.866 & 0.843 & 0.869 \\
\hline
B1 & 0.852 & \textit{0.889} & 0.870 & 0.877 & 0.861 & \textit{0.912} & 0.858 & \textbf{0.889} & 0.907 \\
\hline
C1 & 0.822 & 0.875 & \textbf{0.903} & 0.893 & \textbf{0.910} & 0.889 & 0.828 & 0.865 & \textit{0.920} \\
\hline
A2 & \textbf{0.881} & 0.874 & 0.884 & \textbf{0.903} & 0.888 & 0.888 & 0.867 & 0.861 & 0.898 \\
\hline
C2 & 0.677 & 0.764 & \textit{0.900} & 0.688 & \textit{0.909} & 0.794 & 0.656 & 0.731 & \textbf{0.922} \\
\hline
1M & \textit{0.879} & \textbf{0.890} & \textit{0.900} & \textit{0.902} & 0.907 & \textbf{0.914} & \textbf{0.882} & \textit{0.876} & 0.919 \\
\hline
2M & 0.870 & 0.879 & 0.896 & 0.896 & 0.907 & 0.906 & 0.856 & 0.871 & 0.918 \\
\hline
AM & 0.872 & 0.870 & 0.863 & 0.894 & 0.875 & 0.871 & \textit{0.871} & 0.862 & 0.866 \\
\hline
CM & 0.745 & 0.835 & 0.899 & 0.784 & 0.904 & 0.849 & 0.695 & 0.760 & \textbf{0.922} \\
\hline
\end{tabular}
}
\end{table}

\begin{table}
\centering
\caption{Intra vs. cross domain mean and standard deviation of Dice scores between manual and nnUNet-produced segmentations separately for LV and RV for end-systolic (ES) and end-diastolic (ED) frames on domains A1, A2, A3, B1, B2, B3, C1, C2, C3. In intra-scanner, the training and test sets are from the same scanner but do not have to be from the same pathology group. In intra-pathology, the training and test sets are from the same pathology group but do not have to be from the same scanner. p-values for Mann-Whitney U tests are given for each intra-vs-cross group.}\label{tab_in_cross_dices}
\resizebox{\columnwidth}{!}{
\begin{tabular}{|c|c|c|c|c|}
\hline
  & LV ED & LV ES & RV ED & RV ES \\
\hline
Intra-scanner & 0.944 (0.025) & 0.887 (0.072) & 0.888 (0.057) & 0.838 (0.102) \\
\hline
Cross-scanner & 0.937 (0.030) & 0.873 (0.105) & 0.790 (0.214) & 0.726 (0.253) \\
\hline
In vs cross-scanner p-val & 0.0003 & 0.0285 & 0.0000 & 0.0000 \\
\hline
\hline
Intra-pathology & 0.939 (0.027) & 0.880 (0.090) & 0.837 (0.137) & 0.784 (0.181) \\
\hline
Cross-pathology & 0.940 (0.029) & 0.877 (0.098) & 0.815 (0.208) & 0.751 (0.244) \\
\hline
In vs cross-path. p-val & 0.2716 & 0.4126 & 0.0959 & 0.4778 \\
\hline
\end{tabular}
}
\end{table}

\section{Conclusion}
We have presented an analysis of the effect of domain shift on RV and LV segmentations on a large clinical cine CMR dataset and observed several important results. Firstly, scanner differences seem to play a more important role than pathology group differences on cross-domain segmentation performance and this also means that mixing data from different scanners for training is more helpful than mixing pathologies. Secondly, cross-domain performance is not symmetric, that is, the performance of a model trained on some domain A and tested on some domain B cannot be used to anticipate the performance of a model trained on B and tested on A. This phenomenon was also previously observed in \cite{10.1007/978-3-030-68107-4_24} on the M\&Ms dataset. 
Thirdly, in-scanner vs. cross-scanner performance difference is much larger on RV compared to LV. Lastly, increasing the number of training subjects improves both in-domain and cross-domain performance. The performance improvement is faster for cross-domain on smaller training sets but the cross-domain improvement slows faster than in-domain improvement as training sets get larger. We expect our work to be helpful to researchers studying the effect of domain shift and working on improving generalization of segmentation networks in CMR.

\subsubsection*{Acknowledgements}
This work was funded by the Engineering and Physical Sciences Research Council (EPSRC) programme grant ‘SmartHeart’ \\ (EP/P001009/1) and supported by the Wellcome/EPSRC Centre for Medical Engineering [WT 203148/Z/16/Z]. The research was supported by the National Institute for Health Research (NIHR) Biomedical Research Centre based at Guy's and St Thomas' NHS Foundation Trust and King's College London. The views expressed are those of the authors and not necessarily those of the NHS, the NIHR or the Department of Health.

\bibliographystyle{splncs04}
\bibliography{references}

\begin{thebibliography}{10}
\providecommand{\url}[1]{\texttt{#1}}
\providecommand{\urlprefix}{URL }
\providecommand{\doi}[1]{https://doi.org/#1}

\bibitem{8360453}
Bernard, O., Lalande, A., Zotti, C., Cervenansky, F., Yang, X., Heng, P.A.,
  Cetin, I., Lekadir, K., Camara, O., Gonzalez~Ballester, M.A., Sanroma, G.,
  Napel, S., Petersen, S., Tziritas, G., Grinias, E., Khened, M., Kollerathu,
  V.A., Krishnamurthi, G., Rohe, M.M., Pennec, X., Sermesant, M., Isensee, F.,
  Jager, P., Maier-Hein, K.H., Full, P.M., Wolf, I., Engelhardt, S.,
  Baumgartner, C.F., Koch, L.M., Wolterink, J.M., Isgum, I., Jang, Y., Hong,
  Y., Patravali, J., Jain, S., Humbert, O., Jodoin, P.M.: Deep learning
  techniques for automatic {MRI} cardiac multi-structures segmentation and
  diagnosis: Is the problem solved? IEEE Transactions on Medical Imaging
  \textbf{37}(11),  2514--2525 (2018)

\bibitem{9458279}
Campello, V.M., Gkontra, P., Izquierdo, C., Martin-Isla, C., Sojoudi, A., Full,
  P.M., Maier-Hein, K., Zhang, Y., He, Z., Ma, J., Parreno, M., Albiol, A.,
  Kong, F., Shadden, S.C., Acero, J.C., Sundaresan, V., Saber, M., Elattar, M.,
  Li, H., Menze, B., Khader, F., Haarburger, C., Scannell, C.M., Veta, M.,
  Carscadden, A., Punithakumar, K., Liu, X., Tsaftaris, S.A., Huang, X., Yang,
  X., Li, L., Zhuang, X., Vilades, D., Descalzo, M.L., Guala, A., La~Mura, L.,
  Friedrich, M.G., Garg, R., Lebel, J., Henriques, F., Karakas, M., Cavus, E.,
  Petersen, S.E., Escalera, S., Segui, S., Rodriguez-Palomares, J.F., Lekadir,
  K.: Multi-centre, multi-vendor and multi-disease cardiac segmentation: The
  {M\&M}s challenge. IEEE Transactions on Medical Imaging pp.~1--1 (2021)

\bibitem{doi:10.1056/NEJMc2104626}
Finlayson, S.G., Subbaswamy, A., Singh, K., Bowers, J., Kupke, A., Zittrain,
  J., Kohane, I.S., Saria, S.: The clinician and dataset shift in artificial
  intelligence. New England Journal of Medicine  \textbf{385}(3),  283--286
  (2021)

\bibitem{10.1007/978-3-030-68107-4_24}
Full, P.M., Isensee, F., J{\"a}ger, P.F., Maier-Hein, K.: Studying robustness
  of semantic segmentation under domain shift in cardiac {MRI}. In:
  Puyol~Anton, E., Pop, M., Sermesant, M., Campello, V., Lalande, A., Lekadir,
  K., Suinesiaputra, A., Camara, O., Young, A. (eds.) Statistical Atlases and
  Computational Models of the Heart. M{\&}Ms and EMIDEC Challenges. pp.
  238--249. Springer International Publishing, Cham (2021)

\bibitem{Isensee2021}
Isensee, F., Jaeger, P.F., Kohl, S.A.A., Petersen, J., Maier-Hein, K.H.:
  nnu-net: a self-configuring method for deep learning-based biomedical image
  segmentation. Nature Methods  \textbf{18}(2),  203--211 (Feb 2021)

\bibitem{vonKnobelsdorff-Brenkenhoff2017}
von Knobelsdorff-Brenkenhoff, F., Pilz, G., Schulz-Menger, J.: Representation
  of cardiovascular magnetic resonance in the {AHA / ACC} guidelines. Journal
  of Cardiovascular Magnetic Resonance  \textbf{19}(1), ~70 (Sep 2017)

\bibitem{10.5555/1462129}
Quionero-Candela, J., Sugiyama, M., Schwaighofer, A., Lawrence, N.D.: Dataset
  Shift in Machine Learning. The MIT Press (2009)

\bibitem{UNet}
Ronneberger, O., Fischer, P., Brox, T.: U-net: Convolutional networks for
  biomedical image segmentation. In: Navab, N., Hornegger, J., Wells, W.M.,
  Frangi, A.F. (eds.) Medical Image Computing and Computer-Assisted
  Intervention -- MICCAI 2015. pp. 234--241. Springer International Publishing,
  Cham (2015)

\bibitem{RUIJSINK2020684}
Ruijsink, B., Puyol-Antón, E., Oksuz, I., Sinclair, M., Bai, W., Schnabel,
  J.A., Razavi, R., King, A.P.: Fully automated, quality-controlled cardiac
  analysis from {CMR}: Validation and large-scale application to characterize
  cardiac function. JACC: Cardiovascular Imaging  \textbf{13}(3),  684--695
  (2020)

\bibitem{10.1007/978-3-030-68107-4_10}
Ruijsink, B., Puyol-Ant{\'o}n, E., Li, Y., Bai, W., Kerfoot, E., Razavi, R.,
  King, A.P.: Quality-aware semi-supervised learning for {CMR} segmentation.
  In: Puyol~Anton, E., Pop, M., Sermesant, M., Campello, V., Lalande, A.,
  Lekadir, K., Suinesiaputra, A., Camara, O., Young, A. (eds.) Statistical
  Atlases and Computational Models of the Heart. M{\&}Ms and EMIDEC Challenges.
  pp. 97--107. Springer International Publishing, Cham (2021)

\bibitem{10.1007/978-3-030-68107-4_23}
Scannell, C.M., Chiribiri, A., Veta, M.: Domain-adversarial learning for
  multi-centre, multi-vendor, and multi-disease cardiac {MR} image
  segmentation. In: Puyol~Anton, E., Pop, M., Sermesant, M., Campello, V.,
  Lalande, A., Lekadir, K., Suinesiaputra, A., Camara, O., Young, A. (eds.)
  Statistical Atlases and Computational Models of the Heart. M{\&}Ms and EMIDEC
  Challenges. pp. 228--237. Springer International Publishing, Cham (2021)

\bibitem{10.1007/978-3-030-39074-7_27}
Wang, J., Huang, H., Chen, C., Ma, W., Huang, Y., Ding, X.: Multi-sequence
  cardiac {MR} segmentation with adversarial domain adaptation network. In:
  Pop, M., Sermesant, M., Camara, O., Zhuang, X., Li, S., Young, A., Mansi, T.,
  Suinesiaputra, A. (eds.) Statistical Atlases and Computational Models of the
  Heart. Multi-Sequence CMR Segmentation, CRT-EPiggy and LV Full Quantification
  Challenges. pp. 254--262. Springer International Publishing, Cham (2020)

\end{thebibliography}
\end{document}